\documentclass[a4paper,11pt]{article}
\usepackage[margin=1in]{geometry}
\usepackage{epsfig}
\usepackage{cite}
\usepackage{graphicx}
\usepackage{epstopdf}
\usepackage{color}
\UseRawInputEncoding
\usepackage[nottoc]{tocbibind}
\usepackage[english]{babel}
\usepackage{tabularx}
\usepackage{xparse}
\usepackage{amsmath}
\usepackage{hyperref}
\usepackage{grfext}
\usepackage{amssymb}
\usepackage{caption}
\usepackage{mathtools}
\usepackage{subcaption}
\usepackage{algorithm2e}
\usepackage{float}
\usepackage{subcaption}
\usepackage{multirow}
\restylefloat{table}
\usepackage{booktabs}
\usepackage[dvipsnames]{xcolor}
\usepackage{tikz}
\usepackage{pdflscape}
\usepackage{lscape}

\usepackage{amsmath}
\usepackage{amssymb}
\usepackage{caption}
\usepackage{mathtools}
\usepackage{graphicx}
\usepackage{hyperref}
\usepackage{subcaption}
\usepackage{algorithm2e}
\usepackage{float}
\usepackage[nameinlink]{cleveref}
\usepackage[toc,page]{appendix}
\usepackage{float}
\usepackage{subcaption}
\usepackage{multirow}
\restylefloat{table}
\usepackage{booktabs}
\usepackage[dvipsnames]{xcolor}
\usepackage{tikz}
\usepackage{pdflscape}

\date{\today}
 \normalsize

\newcommand{\bmat}{\left(\begin{array}}
\newcommand{\emat}{\end{array}\right)}
\newcommand{\be}{\begin{equation}}
\newcommand{\ee}{\end{equation}}
\newcommand{\bea}{\begin{eqnarray}}
\newcommand{\eea}{\end{eqnarray}}

\def\gtwid{\mathrel{\raise.3ex\hbox{$>$\kern-.75em\lower1ex\hbox{$\sim$}}}}
\def\ltwid{\mathrel{\raise.3ex\hbox{$<$\kern-.75em\lower1ex\hbox{$\sim$}}}}
\def\gev{{\rm \, Ge\kern-0.125em V}}
\def\tev{{\rm \, Te\kern-0.125em V}}

\def    \be            {\begin{equation}}
\def    \ee            {\end{equation}}
\def    \bea           {\begin{eqnarray}}
\def    \eea           {\end{eqnarray}}
\def\eps{\epsilon}

\def\d{\delta}

\def\nn{\nonumber}
\def\d{\delta}

\begin{document}
\renewcommand{\thefootnote}{\fnsymbol{footnote}}

\vspace{.3cm}

\title{\Large\bf On Warm Natural Inflation and Planck 2018 constraints}

\author
{ \hspace{-3.cm} \it \bf  Mahmoud AlHallak$^{1}$\thanks{mahmoud.halag@unitedschool.ae}, Khalil Kalid Al Said$^{2}$\thanks{khalilium@hotmail.com}, Nidal Chamoun$^{3}$\thanks{nidal.chamoun@hiast.edu.sy}   and Moustafa Sayem El-Daher $^{4,5}$\thanks{ m-saemaldahr@aiu.edu.sy}
 \\\hspace{-3.cm}
 \footnotesize$^1$ Physics Department, Damascus University, Damascus, Syria \\\hspace{-3.cm}
\footnotesize$^2$  Physics Department, University of Doha for Science and Technology, Doha, Qatar \\\hspace{-3.cm}
\footnotesize$^3$ Physics Department, HIAST, P.O. Box 31983, Damascus, Syria \\\hspace{-3.cm}
\footnotesize$^4$ Higher Institute of Laser Applications and Researches, Damascus University, Damascus, Syria \\\hspace{-3.cm}
\footnotesize$^5$ Faculty of Informatics and Communications, Arab International University, Daraa, Syria  %\\\hspace{-3.cm}
}

\date{}
%\date{\today}

\maketitle
%\begin{center}
%\small{\bf Abstract}\\[3mm]
%\end{center}
\begin{abstract}
We investigate Natural Inflation with non-minimal coupling to gravity, characterized either by a quadratic or a periodic term, within the Warm Inflation paradigm during the slow roll stage, in both strong and weak dissipation limits, and show that, in the case of $T$-linearly dependent dissipative term, it can accommodate the spectral index $n_s$ and tensor-to-scalar ratio $r$ observables given by Planck 2018 constraints, albeit with a too small value of the e-folding number to solve the horizon problem, providing thus only a partial solution to Natural Inflation issues. Assuming a $T$-cubically dependent dissipative term can provide a solution to this e-folding number issue.
\end{abstract}

\maketitle

{\bf Keywords}: Warm Inflation, Natural inflation
%\\
%{\bf PACS numbers}:
%\begin{minipage}[h]{14.0cm}
%\end{minipage}

\vskip 0.3cm \hrule \vskip 0.5cm'
%\newpage
%**************************************************************
%**************************************************************

%\begin{document}
%%%%%%%%%%%%%%%%%%%%%%%%%%%%%%%%%%%%%%%%%%
\setcounter{section}{-1} %% Remove this when starting to work on the template.
%\section{How to Use this Template}

\section{Introduction}
Inflationary cosmology \cite{Guth,Linde} is now the dominant perspective in explaining the early universe's physics, solving the flatness, homogeneity\/unwanted relics problems,  and providing a mechanism to interpret the inhomogeneities in the Cosmic Microwave Background Radiation (CMBR). In the standard slow-roll cold inflation models, the universe experiences an exponential expansion, during which density perturbations are created by quantum fluctuations of the inflaton field, followed by the reheating stage, where a temporarily localized mechanism must rapidly distribute sufficient vacuum energy.

Fang and Berera \cite{BereraFang} realized that combining the exponential accelerating expansion phase and the reheating one could resolve disparities assembled by each separately. In \cite{Berera} Berbera proposes a warm inflationary model in which thermal equilibrium is maintained during the inflationary phase and radiation production is started throughout it, i.e., relativistic particles are created during the inflationary period.

Many inflationary models inspired by particle physics, string theory, and quantum gravity have been studied within the context of warm inflation. Visinelli \cite{Visinelli}, derived and analyzed the experimental bounds on warm inflation with a monomial potential, whereas Kamali in \cite{Kamali} investigated the warm scenario with non-minimal coupling (NMC) to gravity with a Higgs-like potential. Warm inflation was constrained by CMB data in \cite{1710.10008}. The authors of \cite{Chamoun_universe} treated the warm scenario with NMC to modified gravity with a special potential motivated by variation of constants. In \cite{Amaek}, warm inflationary models in the context of a general
scalar-tensor theory of gravity were investigated within only the strong limit of dissipation.

The natural inflation (NI) proposed by Freese, Frieman, and Olinto \cite{Freese}, with a cosine potential, is a popular model due to its shift symmetry with a flat potential, preventing significant radiative corrections from being introduced, which gives NI an ability to solve theoretical challenges inherent in slow rolling inflation models.
However, NI is disfavored at greater than $95 \% $ confidence level by current observational constraints from $Planck  2018$ on the scalar-tensor ratio $r$ and spectral index $n_s$ \cite{Planck18, Stein}. Moreover, a more recent analysis of BICEP/Keck XIII in 2018 (BK18) \cite{BK18} has put more stringent bounds on $r$, whereas the authors of \cite{Pappas} discussed a way to amend the discrepancies of NI with data by non-minimally coupling the scalar fields to the Starobinski model ($f(R)$) in the Palatini formalism.  In \cite{Chamoun_JCAP}, it was shown that NMC to gravity within $f(R)$ setting was enough to bring ``cold'' NI to within $95 \% $ confidence levels of the current observational constraints represented by  $Planck 2018$ (TT, EE, TE), BK18 and other experiments (lowE, lensing) separately or combined.

The study of warm NI was pioneered by Visinelli \cite{wni-visinelli} and, separately, by \cite{wni}, then pursued by many others, like \cite{2012.07329}. Applied to primordial black holes (gravitational waves), the setup was analyzed in many articles, say \cite{2207.10394,2105.08045,1910.05238} (\cite{2110.00607,2011.09155}).

The aim of this work is to study NI with NMC to gravity within the Warm paradigm, in both the strong and weak limits of the dissipation term characterizing the warm scenario. We study two forms for the NMC to gravity term, which is generally produced at one-loop order in the interacting theory for a scalar field, even if it is absent at the tree level \cite{Freedman:1974gs}. Actually, in general all terms of the form $(R^i \phi^j, R^{\mu\nu} \partial_\mu\phi \partial_\nu\phi, \ldots)$ are allowed in the action. However, omitting the derivative terms and taking a finite number of loop graphs enforce a polynomial form of the NMC term, and if one imposes CP symmetry on the action the term should include even powers of the inflaton field $\phi$. For simplicity, we include only the quadratic monomial ($\xi \phi^2 R$ ) of dim-4. However, since some microscopic theories may suggest the emergence of an NMC similar in form to the original potential \cite{Salvio}, we also consider an NMC of a periodic form respecting the shift symmetry of the NI potential so to be of the form ($\lambda \left(1+\cos(\frac{\phi}{f})\right)$).

We find that the NI with NMC to gravity within Warm paradigm, in the case of linear dissipative term,  is able to accommodate the $(n_s, r)$ observable constraints, but at a price of getting a small value for the e-folding number $N_e \approx 30$ to solve the horizon and flatness problems. However, one can bring $N_e$ to be acceptable ($\geq 40$), in this $T$-linearly dependent regime, but for $n_s \approx 0.98$, just getting outside the admissibility contours. Studying the case of $T$-cubically dependent dissipative term gave in the strong limit scenario some benchmarks which satisfy the four constraints of $n_s, r, A_s$ and $N_e$ in both cases of quadratic or periodic non-minimal coupling to gravity.

The paper is organized as follows. In section 1, we present the setup of the Warm paradigm, for general potentials, whereas in section 2 we specify the study to NI. In section 3 (4), we study the strong (weak) limit ($Q \equiv \frac{\Gamma}{3 H} >> (<<) 1$) for both quadratic and periodic NMC. In section 5, we study briefly the strong limit scenario when the dissipative term is proportional to the cubic power of the temperature, whereas we end up by conclusions and a summary in section 6.

%%%%%%%%%%%%%%%%%%%%%%%%%%%%%%%%%%%%%%%%%%
\section{Warm Inflation Setup}
\label{warm_section}
\subsection{Arena}
We consider the general local action for a scalar field coupled with radiation and gravity within the Jordan frame,
\begin{equation}
\label{eq:GAction}
S=\int d^4 x \sqrt{-g}\bigg\{\frac{1}{2} \Omega^2(\phi) R + \mathcal{L}_\phi + \mathcal{L}_{\gamma}+\mathcal{L}_{Int} \bigg\}.
\end{equation}
where g is the determinant of the metric $g_{\mu \nu}$,  $ \mathcal{L}_{\gamma}$ is the Lagrangian density of the radiation field and $\mathcal{L}_{Int}$ describes  the  interaction  between  the latter and the inflaton $\phi$ whose Lagrangian density, considered as that of a canonical scalar field, is given by
 \bea
 \mathcal{L}_\phi&=& -\frac{1}{2}g^{\mu \nu }\Delta_\mu \phi  \Delta_\nu \phi-V(\phi),
 \eea
 where $V(\phi)$ is the inflaton potential, whereas $\Omega^2(\phi)$ indicates the NMC between the scalar  field $\phi$ and the gravity described by the usual Einstein-Hilbert action rather than the Starobinski $f(R)$ gravity.

One can take the usual electromagnetic Lagrangian for $\mathcal{L}_{\gamma}$, while we leave aside, for now, the `unknown' interaction density $\mathcal{L}_{Int}$. Carrying out the usual action optimization by changing with respect to metric and approximating the energy-momentum tensor for both the inflaton and the radiation fields by perfect fluids characterized by energy density $\rho$ and pressure $p$, we get the following equation of motion:
\bea
\label{conservation_eq}
\dot{\rho}^{\phi} + 3H \left(\rho^\phi + p^\phi\right) + \frac{1}{2} (\Omega^2)^\prime_\phi \dot{\phi} R + \dot{\rho}^\gamma + 4 H \rho^\gamma &=& 0,
\eea
with $(\Omega^2)^\prime_\phi$ meaning a derivative with respect to $\phi$.
Some remarks are in order here. First, the two terms including the Hubble constant ``$H$" terms, which is known to be related to `total energy' including both those of radiation and inflaton, represents a `direct' coupling between the inflaton
and radiation in contrast to the `indirect' one via the gravity which couples to all fields. Second, $\mathcal{L}_{int}$ contributes an additional  `direct' coupling. However, we still assume that its contribution to the total energy density is negligible, such that
\bea \rho^{tot} &=& \rho^\phi + \rho^\gamma
\eea
There are in the literature some microscopic models for $\mathcal{L}_{int}$ (look for e.g. \cite{Amaek}), but we shall not dwell into their details, but rather assume that its effect is described phenomenologically by a term $\Gamma \dot{\phi}^2$, which can be motivated/justified in a field theory approach specific to the considered microscopic model. As a matter of fact, the $\Gamma$ factor embodies the microscopic physics resulting from the interaction between $\phi$ and other particles, where $\phi$ can usually be assumed to couple to heavy intermediate fields that, in their turn, couple to light radiation fields. As $\phi$ rolls slowly on its potential it triggers the decay of the heavy fields into the light ones generating thus a dissipative term \cite{Ramos,41inRamos}. Another method adopted in warm scenarios is where $\phi$ is a Goldstone boson coupled directly to the light radiation, but gets protected from large thermal corrections due to a symmetry imposed on the model \cite{Ramos,36inRamos}. A supersymmetric model was studied in \cite{ImportanceBeing}, whereas \cite{ZhangDissipative} conceived a model, also supersymmetric, leading to a dissipative factor of the form $\Gamma \propto \frac{T^m}{\phi^{m-1}}$.

We thus assume that these microscopic models lead to a phenomenological term such that:
\bea
\label{friedman_rad_Jordan}
\dot{\rho}^\gamma + 4 H \rho^\gamma &=& \Gamma \dot{\phi}^2
 \eea
whence from Eq.(\ref{conservation_eq}) we have
\bea
\label{friedman_inflaton1_Jordan}
\dot{\rho}^{\phi} + 3H \left(\rho^\phi + p^\phi\right) + \frac{1}{2} (\Omega^2)^\prime_\phi \dot{\phi} R &=& - \Gamma \dot{\phi}^2
 \eea
Using
\bea
\rho^\phi = \frac{1}{2} \dot{\phi}^2 + V &,& p^\phi = \frac{1}{2} \dot{\phi}^2 - V
\eea
we get
\bea
\label{friedman_inflaton2_Jordan}
\ddot{\phi}+ 3H \dot{\phi} V^\prime_\phi + \frac{1}{2} (\Omega^2)^\prime_\phi R &=& - \Gamma \dot{\phi}
 \eea
 Actually, although this `friction' term $\Gamma \dot{\phi}$, describing phenomenologically the decay of $\phi$, may be inadequate to describe the energy transfer from $\phi$ during far out of equilibrium, it is however suitable to describe the energy dissipated by $\phi$ into a thermalized radiation bath \cite{BereraFang}. We shall not discuss the nature of these particles into which $\phi$ decays \cite{Belfiglio,Ford}, rather we shall approximate them by a thermal radiation (namely of photons) such that energy is still dominated by $\phi$, while fluctuations are dominated by thermal, not quantum, ones.

We see that Eq. \ref{conservation_eq} expresses the conservation of total energy, to which one neglects the contribution of $\mathcal{L}_{int}$ which, meanwhile  and through Eqs. (\ref{friedman_rad_Jordan} and \ref{friedman_inflaton1_Jordan}), affects individually both ($\rho^\gamma$ and $\rho^\phi$). There are many possibilities for the dissipative term, but we shall study in this article mainly the case where it depends linearly on temperature ($\Gamma = \Gamma_0 T$), whereas we briefly study in the penultimate section the case of cubical dependence on temperature ($\Gamma = \Gamma_0 T^3$).

During warm inflation, we have $T \gg H$ and due to the inflaton interactions with the matter/radiation, a
bath of particles is continuously produced during the slow roll period, which transits
the universe into a radiation-dominated phase through a smooth transition eliminating, thus, the
need for a reheating stage. Thermal fluctuations dominate over quantum fluctuations, even though $\rho^\gamma$ is neglected versus $\rho^\phi$, which is reflected through the factor
\bea
\label{Q}
Q&=& \frac{\Gamma}{3H} = \frac{\Gamma_0 T}{3 H} \;\;,
\eea
so that the inflation is described to be in the strong (weak) limit regime when $Q \gg 1$ ($Q \ll 1$).

It is convenient to go from Jordan frame to Einstein frame, in which the gravitational sector of the action takes the form of the Hilbert-Einstein action, and the NMC to gravity disappears. Consequently, in Einstein frame, one is able to use the usual equations of general relativity, the inflationary solutions, and the standard slow-roll analysis.

The conformal transformation is defined as:
\bea
\label{conf_trans}
\tilde{g}_{\mu\nu} = \Omega^2(\phi) {g}_{\mu\nu} &\Rightarrow&
\sqrt{-\tilde{g}} = \Omega^4 \sqrt{-g}
\eea
leading to the action expressed in Einstein frame by
\bea
\label{action-einstein}
S &=&\int d^4 x \sqrt{-\tilde{g}}\bigg\{\frac{1}{2}  \tilde{R} -\frac{1}{2} \frac{1+6 ({\Omega}^\prime)^2}{\Omega^2} \tilde{g}^{\mu\nu}\tilde{\nabla}_\mu \phi \tilde{\nabla}_\nu \phi - \frac{V(\phi)}{\Omega^4} \bigg\}+
\int d^4 x \sqrt{-\tilde{g}} \tilde{\mathcal{L}}_\gamma
+S_{Int} . \nonumber \\
\eea
For the radiation field, and since the corresponding integrand in the action is invariant under rescaling, then by Eq. (\ref{conf_trans}) we find that the Lagrangian density (energy-momentum tensor) is divided by $\Omega^4$ ($\Omega^2$), as:
\bea
T_{\mu \nu}^\gamma = \frac{-2} {\sqrt{-g}} \frac{\d\left(\sqrt{-g} \mathcal{L}_\gamma\right)}{\d g^{\mu \nu}}&\rightarrow& \tilde{T}_{\mu \nu}^\gamma = \frac{-2}{\sqrt{-g}} \frac{\d\left(\sqrt{-\tilde{g}} \tilde{\mathcal{L}}_\gamma\right)}{\d \tilde{g}^{\mu \nu}}=\frac{T_{\mu\nu}^\gamma}{\Omega^2} ,\eea
and thus we conclude that the perfect fluid assumption for the radiation field will remain valid in Einstein frame with energy density ($\rho_\gamma = \frac{\rho^\gamma}{\Omega^4}$) and pressure ($p_\gamma = \frac{p^\gamma}{\Omega^4}$) .
Taking the definition of temperature:
\bea
\label{tempe}
\rho_{(\gamma)}^{\gamma} = C_\gamma {T_{(\gamma)}^{\gamma}}^4 &:& C_\gamma=\frac{\pi^2 g_*}{30},
\eea with $g_*$ denoting the number of created massless modes, we see that the temperature scales by $1/\Omega$ going from Jordan to Einstein frame:
\bea
\label{T-transform}
T \xrightarrow{\mbox{Jordan}\rightarrow \mbox{Einstein}} T/\Omega
\eea

As to the inflaton and gravity sector, we see that in Einstein frame there is a `pure' GR gravity part, whereas we have a non-canonical kinetic term for the inflaton scalar field, which can be put in a canonical form by defining a new field $\chi$,  related to  $\phi$ by:
\begin{equation}
\label{Z_metric}
\bigg(\frac{d\phi}{d \chi} \bigg)^2 \equiv \frac{1}{Z^2} = \frac{\Omega^2}{1+6 ({\Omega}^\prime_\phi)^2} = \frac{2 \Omega^4}{2 \Omega^2 + 3 (({\Omega^2})^\prime_\phi)^2},
\end{equation} so to get (from now on, we drop the tilde off, but we keep in mind that all calculations are carried out in Einstein frame):
\bea
\label{action-einstein}
S &=&\int d^4 x \sqrt{-g}\bigg\{\frac{1}{2}  R -\frac{1}{2} g^{\mu\nu}\nabla_\mu \chi \nabla_\nu \chi - U(\chi) \bigg\}+
S_\gamma
+S_{Int},
\eea
where
\bea
U(\chi) &=& \frac{V(\phi(\chi))}{\Omega^4}
\eea

A spatially flat Friedmann-Robertson-Walker (FRW) Universe gives the energy density $\rho_\chi$ and the pressure $ p_\chi$ of the inflaton field as,
\bea
\label{rho-p_chi}
\rho_\chi = \frac{1}{2} \dot{\chi}^2 + U(\chi) &,&
p_\chi = \frac{1}{2} \dot{\chi}^2 - U(\chi)
\eea
with Friedman equation given by,
\begin{equation}
\label{friendmn_1}
H^2 = \frac{1}{3} \rho_{tot} = \frac{1}{3}(\rho_\chi+\rho_\gamma).
\end{equation}

For the interaction Lagrangian $\mathcal{L}_{int}$, and lacking a model-independent Lagrangian term leading to the RHS of (\ref{friedman_inflaton2_Jordan}), we shall argue by comparison to the cold inflation scenario in order to find the corresponding equation in Einstein frame. Note that, unlike standard studies (\cite{Kamali}) where the damping term is introduced in Einstein frame, we espouse the viewpoint that the field approach models justifying the damping term form are to be defined in the original Jordan frame. However, we shall show that under an approximation, which we shall adopt, the form would be similar in the two frames.
Actually, the Hubble parameter transformation  has an inhomogenous term \cite{fujii}\footnote{Note however that the ``measurable'' Hubble parameter in Einstein frame will be the one corresponding to dropping the inhomogeneous term \cite{Yamaguchi}.  }:
\bea
\label{H-transform}
H \xrightarrow{\mbox{Jordan}\rightarrow \mbox{Einstein}} H/\Omega + \frac{\dot{\overset{\Huge\frown}{\log \Omega}}}{\Omega}
\eea
then looking at Eq. (\ref{T-transform}) and dropping/neglecting the inhomogeneous logarithmic variation of $\Omega$, we see that $\frac{T}{H}$ is conformally invariant, and likewise the factor  $Q$ (Eq. \ref{Q}) is also invariant. In Einstein frame, the field will undergo slow rolling generating inflation where one assumes approximate constancy for both $H$ and $T$ \cite{sciencedirect}, so one can consider $Q$ as constant in Einstein frame, and thus also in Jordan, frame. We know that in cold inflation, including NMC to gravity, a Jordan-frame Euler-Lagrange-type equation expressing metric stationarity:
\bea
\label{EL_NMC_Jordan}
\ddot{\psi}_J+ 3H_J \dot{\psi}_J + (V_J)^\prime_{\psi_J} + \frac{1}{2} (\Omega^2)^\prime_{\psi_J} R &=& 0
 \eea would lead in Einstein frame to a standard GR inflationary equation:
\bea
\label{friedman_Einstein}
\ddot{\psi}_E+ 3H_E \dot{\psi}_E + (V_E)^\prime_{\psi_E} &=& 0
 \eea
We see now that using Eq. (\ref{Q}) in Eq. (\ref{friedman_inflaton2_Jordan}),
we get an equation similar to Eq. (\ref{EL_NMC_Jordan}), but with $(3 H_J)$ replaced by $(3(1+Q)H_J)$, where $Q$ is approximately constant, then we conclude that we get in Einstein frame an equation similar to Eq. (\ref{friedman_Einstein}) with ($H_E$) replaced by ($H_E (1+Q)$). Rewriting $Q$ in Einstein frame we get (dropping the subscript E) in Einstein frame:
\bea
\label{chi-einstein}
\ddot{\chi}+3 H \dot{\chi} +U^\prime_\chi &=& - \Gamma \dot{\chi} = -\Gamma_0 T \dot{\chi}
\eea
So, the upshot here is that we can use, under some approximation and for a damping factor linearly proportional to temperature, the above standard form, albeit starting from a free parameter $\Gamma_0$ defined originally in Jordan frame. By conservation of energy, We get:
\bea
\label{gamma-einstein}
\dot{\rho}_\gamma + 3 H (\rho_\gamma + p_\gamma) = \dot{\rho}_\gamma + 4 H \rho_\gamma  &=& + \Gamma \dot{\chi}^2
\eea
The fundamental equations for warm inflation within the slow roll approximation ($\dot{\chi}^2 \ll U, \ddot{\chi} \ll \dot{\chi}, \dot{\rho}_\gamma \ll \rho_\gamma$) are:
\bea
\label{warm_inflation_equations}
H^2 \approx U/3 &,& \dot{H} \approx -\frac{1}{2} (1+Q) \dot{\chi}^2 , \\
\rho_\gamma \approx \frac{3}{4} Q \dot{\chi}^2 &,& \dot{\chi} \approx -\frac{U^\prime_\chi}{3 H (1+Q)}.
\eea
Using Eq. (\ref{tempe}), we get
\bea
\label{temperature}
T&=& \left(\frac{1}{4 C_\gamma} \frac{Q}{(1+Q)^2} \frac{(U^\prime_\chi)^2}{U}\right)^{\frac{1}{4}}
\eea

\subsection{Power spectrum}
We define the following slow roll parameters:
\bea
\label{slowroll parameters}
\eps &=& \frac{1}{2}\left(\frac{U^\prime_\chi}{U}\right)^2 = Z^2 \eps^\phi, \\
\eta &=& \frac{U^{\prime\prime}_{\chi\chi}}{U} = Z^2 \eta^\phi + Z Z^\prime_\phi \sqrt{2 \eps^\phi}, \\
\beta &=& \frac{\Gamma^\prime_\chi U^\prime_\chi}{\Gamma U} = Z^2 \beta^\phi ,
\eea
where $\eps^\phi, \eta^\phi, \beta^\phi$ correspond to the same definitions with the derivative carried out with respect to the field $\phi$. One can show that the slow roll regime is met provided we have
\bea
\label{slow roll condition}
\eps, \eta, \beta &\ll& 1+Q
\eea
The spectrum of the adiabatic density perturbations generated during inflation is given by \cite{Kamali} (the star * parameter denotes parameter at horizon crossing):
\bea
\label{scalar_power_spectrum}
\Delta_R(k) &=& A_s \left(\frac{k}{k_*}\right)^{n_s(k)-1}= P_0(k/k_*) \mathcal{F}(k/k_*) : \\
P_0(k/k_*) = \left( \frac{H_*^2}{2\pi \dot{\chi}_k}\right)^2 &,&  \mathcal{F}(k/k_*) = \left( 1+2 \nu_k + \omega_k \right) G(Q_*): \\
\nu_k = \frac{1}{e^{\frac{H}{T}}-1} &,& \omega_k = \frac{T}{H} \frac{2\sqrt{3} \pi Q_k}{\sqrt{3+4\pi Q_k}}  \label{generalomega},\\
A_s &=& \Delta_R(k_*)=\left( \frac{H_*^2}{2\pi \dot{\chi}_{k_*}}\right)^2 \left( 1+2 \nu_{k_*} + \omega_{k_*} \right) G(Q_*)
\label{A_s},
\eea
where $A_s$ represents the amplitude of the CMB fluctuations at the scale $k_*$, and where the modification function $G$, which is due to coupling between the inflaton field and radiation fluctuations, is given numerically for a linearly $T$-dependent dissipation by:
\bea
\label{G}
G(Q) &=& 1+0.335 Q^{1.364} + 0.0815 Q^{2.315}.
\eea

The curvature perturbation spectrum has been measured by PLANCK (WMAP) at
$68\%$ Confidence Level at the fixed wave number $k_{\star} = 0.05(0.002) \mbox{ Mpc}^{-1}$ as
\cite{planck2013}(\cite{wmap2013})
\bea
\label{A_s_Constraints}
A_s \in [2.136,2.247] \times 10^{-9} &,& \left( \in [2.349,2.541] \times 10^{-9}\right)
\eea
and thus the model seeks to reproduce these observational constraints, or at least to reproduce their order of magnitude ($10^{-9}$).

We see that the cold inflation is restored when ($\nu_k$), the Bose-Einstein distribution in a radiation bath of temperature $T$, and $\omega_k$, due to thermal effects, both go to zero:
\bea
\left( \nu_k \rightarrow 0 \,\,\, \& \,\,\, \omega_k \rightarrow  0 \right) &\Rightarrow& \mbox{cold inflation}
\eea
The observable spectral index is given by:
\bea
\label{n_s}
n_s-1 &=& \left.\frac{d \log \Delta_R(k)}{d \log k}\right|_{k=k_*} = \frac{1}{H}\frac{d \log \Delta_R(k)}{dt} = \frac{1}{H \Delta_R}\frac{d\Delta_R}{dt}
\eea
whereas the observable $r$, the tensor-to-scalar ratio, is given by:
\bea
\label{r}
r = \frac{\Delta_T(k)}{\Delta_R} = \frac{2 H^2}{\pi^2 \Delta_R} = \frac{16 \eps}{(1+Q)^2} \mathcal{F}^{-1}
\eea
We distinguish two limit regimes.
\begin{itemize}
\item {\bf Strong Limit $Q \gg 1$}:

Here, using Eq. (\ref{temperature}), one can show that
\bea
\label{strongT}
T&=& \left(\frac{Z^2 (U^\prime_\phi)^2}{4 H C_\gamma \Gamma_0}\right)^{1/5}
\eea
We have via Eq. (\ref{generalomega}):
\bea
\label{strongomega}
\omega_k &=& T \sqrt{\frac{\pi \Gamma}{H^3}} = \frac{T}{H}\sqrt{3 \pi Q}
\eea
Thus, $1+\nu_k \approx \frac{T}{H} \ll \omega_k$, and one gets:
\bea
\label{strong_delta_R}
\Delta_R = \Delta R_s \; G&:& \Delta R_s=\frac{\sqrt{3}TH}{8\sqrt{\pi}} \eps^2 Q^{\frac{5}{2}}
\eea
Thus we get
\bea
n_s-1 &=&\frac{1}{H\Delta R_s} \frac{d\Delta R_s}{dt} + \frac{\dot{Q}}{H} \frac{G^\prime_Q}{G}
\eea
The first term will give after lengthy calculations (look at \cite{Visinelli}:
\bea
\frac{1}{H\Delta R_s} \frac{d\Delta R_s}{dt} &=& \frac{1}{Q} \left( -\frac{9}{4} \eps + \frac{3}{2} \eta - \frac{9}{4} \beta \right)
\eea
whereas we get for the second term:
\bea
 \frac{\dot{Q}}{H} \frac{G^\prime_Q}{G} &=& \frac{2.315}{Q} (\eps-\beta)
\eea
where we have used the identity:
\bea
\label{identity}
\frac{\dot{Q}}{HQ} = \frac{\dot{\Gamma}}{H \Gamma} - \frac{\dot{H}}{H^2} &=& \frac{-1}{1+Q} (\beta -\eps)
\eea
Thus we get:
\bea
\label{Strong_ns}
n_s-1 &=& \frac{1}{Q} \left( -\frac{9}{4} \eps + \frac{3}{2} \eta - \frac{9}{4} \beta +2.3 \eps - 2.3 \beta\right)
\eea
Note that $n_s$ involves the temperature $T$ through the expression of $Q=\frac{\Gamma_0 T}{3 H}$. Also, the temperature $T$  plays a role in determining the ``end of inflation'' field ($\phi_f$) being the argument of the slow roll parameter ($\eps, \eta, \beta$) when it equals $1+Q=1+\frac{\Gamma_0 T}{3H}$, whichever amidst the three meets the equality first. Determining $\phi_f$ allows to compute the e-folding number by:
\bea
\label{efolding}
N_e \equiv \log \frac{a_{\mbox{end}}}{a_k} = \int_t^{t_f} H dt = \int_{\chi_k}^{\chi_f} H \frac{d\chi}{\dot{\chi}} \approx \int^{\chi_k}_{\chi_f} \frac{U}{U^\prime_\chi} (1+Q) d\chi
 = \int^{\phi_k}_{\phi_f} \frac{U}{U^\prime_\phi} (1+Q) Z^2 d\phi
 \eea
The initial time when the inflation started is taken to correspond to the horizon crossing when the dominant quantum fluctuations freeze transforming into classical perturbations with observed power spectrum.

As for the tensor-to-scalar ratio we get
\bea
\label{Strong-r}
r &=& \frac{H}{T}\frac{16 \eps}{Q^{5/2}} G^{-1} =  \frac{H}{T}\frac{16 \eps}{0.0185 Q^{4.815}}
\eea

\item{\bf Weak Limit $Q \ll 1$}

Using Eq. (\ref{temperature}), one can show that
\bea
\label{weakT}
T&=& \left(\frac{Z^2 (U^\prime_\phi)^2 \Gamma_0}{36 C_\gamma H^3}\right)^{1/3}
\eea
From eq. (\ref{generalomega}), we have
\bea
\label{weakomega}
\omega_k &=&  \frac{2 \pi \Gamma T}{3 H^2} = \frac{2 \pi T Q}{H}
\eea
Thus, $1+\nu_k \approx \frac{T}{H} \gg \omega_k$, and one gets:
\bea
\label{weak_delta_R}
\Delta_R = \Delta R_w \; G&:& \Delta R_w=\frac{4 T H}{\pi} \eps^2
\eea
Thus we get
\bea
n_s-1 &=&\frac{1}{H\Delta R_w} \frac{d\Delta R_w}{dt} + \frac{\dot{Q}}{H} \frac{G^\prime_Q}{G}
\eea
The first term will give after lengthy calculations (look at \cite{Visinelli}:
\bea
\frac{1}{H\Delta R_w} \frac{d\Delta R_w}{dt} &=& 1-6 \eps + 2 \eta + \frac{\omega_k}{1+\omega_k} \left( \frac{15 \eps -2 \eta -9 \beta}{4}\right)
\eea
which gives, under the condition:
\bea
\label{condition}
\omega_k=\frac{2 \pi T Q}{H} &\ll& 1,
\eea
the answer
\bea
\frac{1}{H\Delta R_w} \frac{d\Delta R_w}{dt} &=& 1-6 \eps + 2 \eta + \frac{2 \pi \Gamma_0 T^2}{3H^2} \left( \frac{15 \eps -2 \eta -9 \beta}{4}\right)
\eea
As to the second term, we get using Eq. (\ref{identity})
\bea
 \frac{\dot{Q}}{H} \frac{G^\prime_Q}{G} &=& 0.456 Q^{1.364} (\eps-\beta)
\eea
Thus we get:
\bea
\label{weak_ns}
n_s-1 &=& 1-6 \eps + 2 \eta + \frac{2 \pi \Gamma_0 T^2}{12 H^2} (15 \eps -2 \eta -9 \beta) + 0.456 Q^{1.364} (\eps-\beta)
\eea
As for the tensor-to-scalar ratio we get, using $G \approx 1$, the following
\bea
\label{Strong-r}
r &=& \frac{16 \eps}{(1+Q)^2} \mathcal{F}^{-1} =  \frac{8 H \eps}{T}= \frac{8 H Z^2 \eps^\phi}{T}
\eea

\end{itemize}

\section{Natural Inflation}
The potential in the NI is periodic of the form
\bea
\label{NI-potential}
V &=& V_0 \left( 1+ \cos(\frac{\phi}{f})\right)
\eea
where $V_0$ is a scale of an effective field theory generating this potential, and $f$ is a symmetry
breaking scale. As mentioned in the introduction we shall consider two well motivated forms of NMC to gravity:
\begin{itemize}
\item{Quadratic NMC}:
\bea
\label{NMC-quadratic}
\Omega^2(\phi) &=& 1+ \xi \phi^2
\eea
which is considered a leading order of terms allowed in the action generated by loops in the interacting theory. $\xi$ is the free parameter coupling constant characterizing the strength of the NMC to gravity.

\item{Periodic NMC}
\bea
\label{NMC-periodic}
\Omega^2(\phi) &=& 1+ \lambda \left(1+\cos(\frac{\phi}{f})\right)
\eea
which is similar in form to the original potential, allowing it to be justified in some microscopic models.

\end{itemize}

It is well known that Cold Natural inflation with NMC is not enough to accommodate data. In \cite{Chamoun_JCAP}, we showed that Cold Natural inflation with NMC and $F(R)$-modified gravity
was viable. Here we are trying to dispense of the modification of gravity ingredient, while assuming, instead, the Warm scenario. We shall see that $2$ constraints out of $3$ can be met for the Warm NI with NMC.

The strategy would amount to carry out an exhaustive scanning of the free parameters space (that of $\phi_*, \Gamma_0, f, V_0, \xi \mbox{ or } \lambda$) and compute for each `benchmark' the corresponding $n_s, r$ and $\phi_f$, the latter making one of the slow parameters equal to $(1+Q)$, which would allow us to compute the e-folding number $N_e$, which with ($n_s, r$) would constitute the observational constraints to be accommodated.  As to the number of relativistic degrees of freedom of radiation,  we use $g_*(T) = 228.75$, i.e. $C_\gamma=75.2557$, corresponding to the number of relativistic degrees of freedom in
the minimal supersymmetric standard model at temperatures greater than the electroweak
phase transition.

\section{Comparison to Data: Strong case  }
We carried out an extensive scan over the free parameters space, and for each point we computed $(n_s, r)$ and $N_e$. We could not find benchmarks meeting the constraints of ($n_s, r$) at $95\%$ confidence levels according to the 2018 Planck (TT, EE, TE), BK18 and other experiments (lowE, lensing) separately or combined, which would allow also for acceptable $N_e \geq 40$ in order to solve the flatness and horizon problems. Accommodating $(N_e=40, r)$ was possible but at the expense of getting ($n_s$) a bit large.
\vspace{0.3cm}
\begin{itemize}

\item {Quadratic NMC}
\begin{figure}[H]
\includegraphics[width=11.5cm]{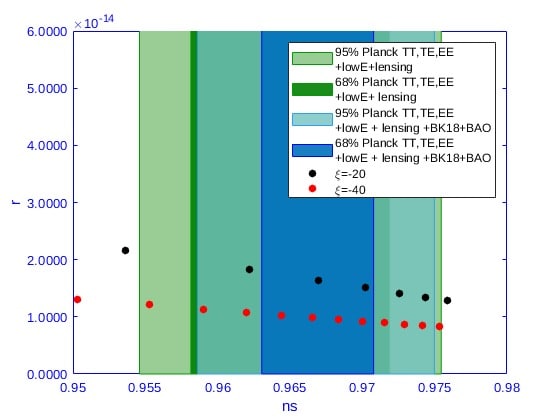}
\caption{ Predictions of warm natural inflation with Quadratic NMC to gravity in the Strong limit. We took the values in units where Planck mass is unity ($\Gamma_0=7000, f=5, V_0=5 \times 10^{-6}$). For the black (red) dots, we have $\xi=-20 (-40), \phi_* \in [3\times 10^{-4},0.0015]$ ($\in[3 \times 10^{-4}, 0.0029]$) corresponding to $N_e \in [14.8,30.4] (\in [9.3,26.4])$. $Q$ in both cases is of order $10^3$ \label{fig1}}
\end{figure}

Fig. (\ref{fig1}) shows the results of scanning the parameters space in the case of Strong limit Warm NI with Quadratic NMC to gravity. One could accommodate ($n_s, r$) but with too little $N_e$. In the figures, the two colors dots  correspond to two choices of the coupling $\xi=-20, -40$.

Looking to meet the e-folds constraint, we imposed ($N_e=40$) with ($\xi=-20$), and fixed the values of ($\Gamma_0, f, V_0$) as before, while scanned over $\phi_*$. We found the `bench mark':
($\phi_*=0.0029$) giving the required e-folds with $r=1.03 \times 10^{-14}$ and $Q$ of order $1.3 \times 10^3$. However, the scalar spectral index $n_s$ was large ($n_s = 0.98$) outside the acceptable contours.
\vspace{0.3cm}

\item{Periodic NMC}
\begin{figure}[H]
\includegraphics[width=11.5cm]{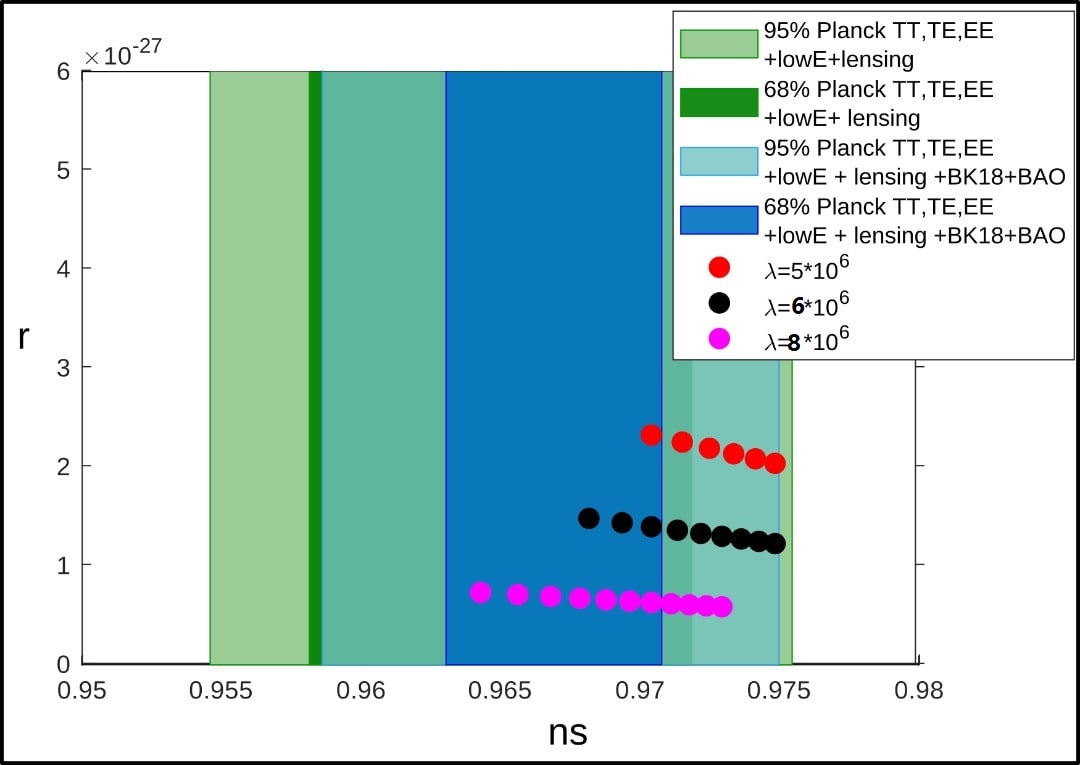}
\caption{Predictions of warm natural inflation with Periodic NMC to gravity in the Strong limit. We took the values in units where Planck mass is unity ($\Gamma_0=1000, f=100, V_0=1 \times 10^{-6}$). For the red (black, pink)  dots, we have $\lambda=5 \times 10^6 (6 \times 10^6, 8 \times 10^6), \phi_* \in [1\times 10^{-4},1.5 \times 10^{-4}]$ ($\in [1\times 10^{-4},1.8 \times 10^{-4}], \in [1\times 10^{-4},2 \times 10^{-4}] $) corresponding to $N_e \in [24.5,29.3] (\in [22.6,29.3], \in [19.9, 21.6])$. $Q$ in all cases is of order $6 \times 10^3$  \label{fig2}}
\end{figure}

Fig. (\ref{fig2}) shows the results of scanning the parameters space in the case of Strong limit Warm NI with periodic NMC to gravity. As in the case of Quadratic NMC, one could accommodate ($n_s, r$) but with too little $N_e$. In the figures, the three colors dots  correspond to three choices of the coupling $\lambda (\times 10^{-6}) =5, 6, 8$.

Again, one could meet the acceptable value ($N_e=40$) with ($\lambda = 5 \times 10^{-6}$) and the values of ($\Gamma_0, f, V_0$) as before,  through scanning over $\phi_*$, and finding a `bench mark':
($\phi_*=5 \times 10^{-4}$) giving the required e-folds ($N_e=40.14$) with $r=4.2 \times 10^{-20}$ and $Q$ of order $1.3 \times 10^4$. However, the scalar spectral index $n_s$ was again large ($n_s = 0.98$) outside the acceptable contours.
\end{itemize}

\section{Comparison to Data: Weak case  }
As in the case of Strong limit, we performed an exhaustive scan over the free parameters, and for each point we computed $(n_s, r)$ and $N_e$. Again, the search was negative for benchmarks meeting the constraints of ($n_s, r$) at $95\%$ confidence levels of the Planck 2018 data, with acceptable $N_e \geq 40$. Unlike the strong limit, we could not accommodate $(N_e=40)$ even with out-of-range ($n_s, r$).
\begin{figure}[H]
\includegraphics[width=11.5cm]{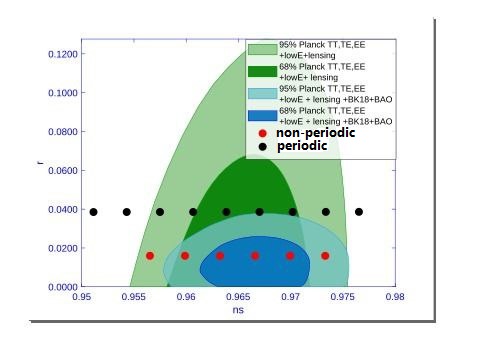}
\caption{ Predictions of warm natural inflation with NMC to gravity in the Weak limit. For the Quadratic (Periodic) NMC  in red (black) dots, we took (the values are given in units where Planck mass is unity): $\Gamma_0=7.14 \times 10^{-7}, f=2, V_0=2.25 \times 10^{-15}$. We fixed $\phi_* = 2 (6.9)$ and scanned over $\xi (\lambda)$  $\in [1.99, 2.00] ([1.04, 1.06])$. We found $n_s \in [0.95, 0.97] ([0.95, 0.97])$, $r \in [0.015, 0.016] ([0.0385, 0.0386])$, and we got $N_e \approx 0.96 (0.27)$. In both cases of NMC we had $Q$ of order $10^{-4}$.
\label{fig3}}
\end{figure}
Fig. {\ref{fig3}} shows some results of our scan. In both cases of Quadratic and Periodic NMC to gravity we took $\Gamma_0=7.14 \times 10^{-7}, f=2, V_0=2.25 \times 10^{-15}$. The dots correspond to fixing the horizon crossing field and scanning over the NMC coupling ($\xi, \lambda$). As the figure shows, even though one could accommodate the observables ($n_s, r$), however the e-folds number was always too small to be acceptable, which means the ingredient of ``warm scenario' was not enough to solve the problems of the NI with NMC.
%%%%%%%%%%%%%%%%%%%%%%%%%%%
\section{Cubic Dissipative Term}
In order to tackle the ``insufficient $N_e$" problem, which is fatal for any plausible inflationary model, we consider the case of cubic dissipation factor ($\Gamma = \Gamma_0 T^3$).

As said earlier, some microscopic models may lead to the $T$-cubically dependent dissipation factor describing a decay of $\phi$ into radiation fields through intermediate heavy fields. However, we shall not suppose this form in the original Jordan frame, but rather assume it directly in Einstein frame and investigate the results.

The analytical expressions of the resulting ($n_s, r$ and $N_e$) are too cumbersome to be stated here. However, for the strong limit cubicly $T$-dependent dissipation, one can approximate the CMB fluctuations amplitude $A_s$ by \cite{Visinelli}:
\bea
\label{AsStronglimit}
A_s &=& \frac{1}{8 \pi} \left(\frac{9}{2 \pi^2 C_\gamma}\right)^{1/4} \left( \frac{Q^3 U}{\eps_k}\right)^{3/4} G(Q)
\eea
where, in contrast to the linearly $T$-dependent dissipation case (Eq. \ref{G}), the modification function $G$ is given now by \cite{Benetti}:
\bea
\label{Gcubic}
G(Q) &=& 1+4.981 Q^{1.946} + 0.127 Q^{4.330}.
\eea

We scanned numerically over five free parameters $(f, V_0, \Gamma_0, \phi_*)$ and $\xi (\lambda)$ in the quadratic (periodic) NMC scenario, and for each point in the parameter space we computed $\phi_{end}$, corresponding to one of the slow rolling parameters being equal to unity, then obtained $(n_s, r, A_s)$, to check they meet the experimental observations, with a suitable computable $N_e$. We evaluate finally $(Q, T, H)$ to check that the conditions of strong limit regime ($Q \gg 1$) and warm inflation scenario ($T/H > 1$) are satisfied. We found the following benchmark intervals:

\begin{itemize}
  \item Non-periodic NMC:

  Scanning over ($\xi \in [19,22]$) with
  \bea
  &\phi_*= 0.063, f=5, \Gamma_0=3 \times 10^{-9} , V_0 = 2\times 10^{-21},
  \eea
  we found acceptable points with the following ranges:
  \bea
  &\phi_{end} \in [8.10, 8.34], A_s \in [2.085, 2.128] \times 10^{-9},& \nn \\
  & n_s \in [0.9433, 0.9864], r \in [1.86, 1.93] \times 10^{-13},&\nn \\ & N_e \in [73.47, 73.50], Q \in [28.48, 28.73] ,& \nn \\
  &T \in [9.8818,9.8895] \times 10^{-7}, H \in [3.3582, 3.3954]\times 10^{-11}&
  \eea
  with $N_e, A_s, Q$ ($\phi_{end}, n_s, r, T, H$) increasing (decreasing) with $\xi$.

  \item Periodic NMC:
  Scanning over ($\lambda \in [25,73]$) with
  \bea
  &\phi_*= 9, f=5, \Gamma_0= 10^{10} , V_0 = 1 \times 10^{-18},
  \eea
  we found acceptable points with the following ranges:
  \bea
  &\phi_{end} \in [0.1267, 0.3601], A_s \in [0.095,1.5] \times 10^{-9},  &\nn \\
  & n_s \in [0.9519, 0.9691], r \in [1.4,2.86] \times 10^{-13},&\nn \\ & N_e \in [41, 77], Q \in [20.47, 23.72], & \nn \\ &
  T \in [3.7,5.6]\times 10^{-7}, H \in [0.884,2.49] \times 10^{-11}
  \eea
  with $\phi_{end}, r$ ($n_s, N_e, A_s, Q, T, H$) increasing (decreasing) with $\lambda$.
\end{itemize}

Note also that only `mild' constraints on Hubble parameter values $\in [10^{-24},10^{14}]$ GeV (or $\in [10^{-43},10^{-5}]$ in natural units) exist in the literature \cite{Jiang}, which are respected in the above values for both quadratic and periodic NMC cases.

\begin{figure}
\hspace{-3cm}
 \includegraphics[width=16.cm]{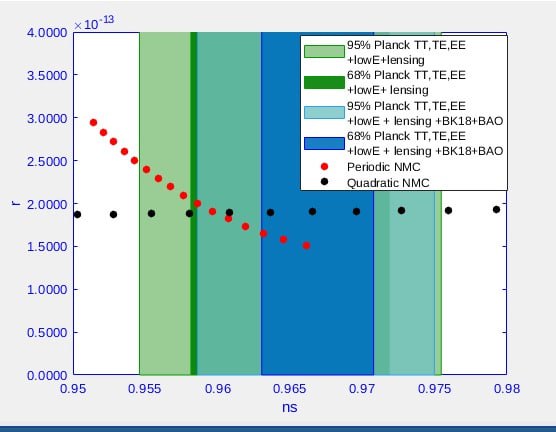}
  \begin{tabular}{cc}
  \hspace{-3cm}
   \includegraphics[width=8.cm]{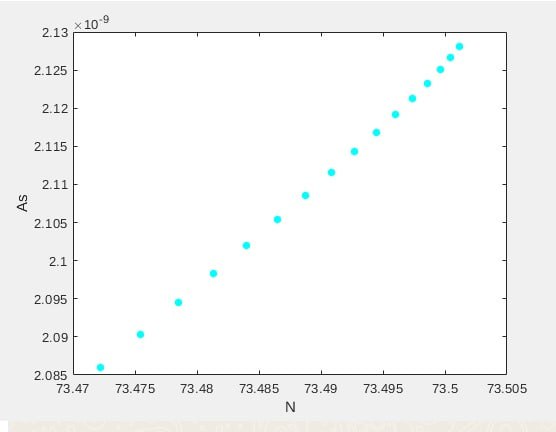}
   &
   \includegraphics[width=8.cm]{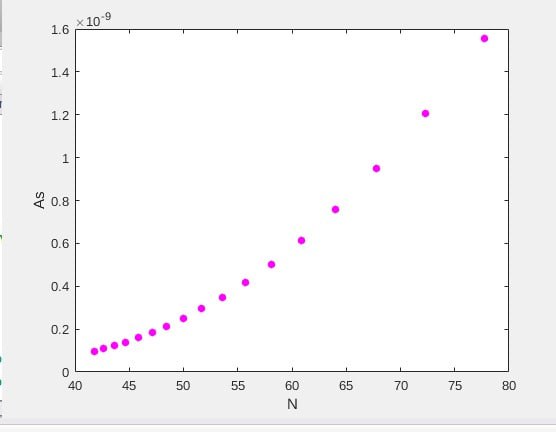}\\
   (a)&(b)
  \end{tabular}
    \caption{ Above: Cubic Dissipation Case for both Quadratic and Periodic NMC, showing consistency with observational data regarding ($n_s, r$). Bottom: The quadratic NMC (a) accommodates ($A_s, N_e$), whereas the periodic (b) NMC accommodates $N_e$ and the magnitude order of $A_s$. }
  \label{FigCubicDissipation}
\end{figure}

Fig. (\ref{FigCubicDissipation}) shows the mentioned benchmark acceptable points, and also shows the allure of $A_s$ with respect to the e-foldings number $N_e$ in the case of quadratic (a) and periodic (b) NMC. The $A_s$-values are acceptable in the quadratic case. However, for the periodic case, we see that the order of magnitude ($10^{-9}$) for $A_s$ is reproduced, albeit with prefactors not reaching the constraints of (Eq. \ref{A_s_Constraints}), unless at the expense of large e-foldings number values. We do not consider large $N_e > 80$ as an exclusionary sign, since there are many inflationary models arguing for such high values of `total' $N_e$ not contradicting `observable' constaints of $N_e$ (\cite{Liddle}).

%%%%%%%%%%%%%%%%%%%%%%%%%%%%%%%%%%%%%%%%%%
\section{Summary  and Conclusion}
We discussed in this paper the scenario of warm NI with NMC to gravity. It is well known that NI with NMC and modified gravity is viable considering the Planck 2018 data. We kept the GR Einstein-Hilbert action and examined the possibility of whether assuming the 'warm' paradigm could make the NI with NMC viable. Within the warm paradigm, we introduced the `phenomenological' damping factor in Jordan frame, and examined the approximation which would put it in the same form in Einstein frame. We restricted first our study to the case where the damping constant is linearly proportional to temperature.

We found that in the strong limit, the model is able to accommodate the spectral observables ($n_s, r$) but with a small e-fold number reaching $N_e \sim 30$. However, the points allowing for larger $N_e \geq 40$ would lead to spectral observables slightly out of range.

In the weak limit, the allowed parameter space for ($n_s, r$) is far narrower than in the strong limit, but the corresponding $N_e$ is too small ($N_e \leq 1$) to be remedied even at the price of pushing ($n_s, r$) considerably out of range.

We, second, treated briefly the case of the damping constant being proportional to the temperature raised to the power three. We, upon scanning the free parameters, found some benchmark points, in the limit of strong $Q$, satisfying the four constraints on ($n_s, r, A_s$ and $N_e$).

We conclude that the `warm' ingredient may be enough to solve the problems of NI, provided one explores different forms of $\Gamma$-dependence on $T$. Alternatively, a  possible combination of `warm' paradigm plus other mechanism, such as assuming Palatini formalism rather than the metric one, may be fruitful if one wants to make a warm NI with NMC viable.

\vspace{6pt}

{\bf Acknowledgments:} N. Chamoun acknowledges support from ICTP-Associate program (Italy), from the Alexander von Humboldt Foundation (Germany), and from the PIFI program at the Chinese Academy of Sciences. M. A., N.C. and M.S.E.-D. thank the President of Damascus University, Prof. Muhammad Osama AlJabban, for his help and support.

\end{document}